\documentclass[structabstract]{aa}  
\usepackage{graphicx}
\usepackage{natbib}
\usepackage{txfonts}
\usepackage{multirow}
\usepackage{aalongtable}
\begin{document}
   \title{TIMASSS : The IRAS16293-2422 Millimeter And Submillimeter Spectral Survey: Tentative Detection of Deuterated Methyl Formate (DCOOCH$_3$)
   \thanks{Based on observations carried out with the IRAM 30 m telescope. IRAM is supported by INSU/CNRS (France), MPG (Germany) and IGN (Spain).},
\thanks{Based on analysis carried out with the CASSIS software and the JPL (http://spec.jpl.nasa.gov/) and CDMS (http://www.ph1.uni-koeln.de/cdms/) spectroscopic databases. CASSIS has been developed by CESR-UPS/CNRS (http://cassis.cesr.fr).}
}

  \author{K. Demyk \inst{1, 2}
         \and
          S. Bottinelli \inst{1, 2} 
                   \and
         E. Caux \inst{1, 2}
          \and
         C. Vastel \inst{1, 2}
         \and
         C. Ceccarelli \inst{3}
          \and
	  C. Kahane \inst{3}
	  \and
	 A. Castets \inst{3}
          }

\institute{CESR (Centre d'\'etude Spatiale des Rayonnements), 
              Universit\'e de Toulouse [UPS], F-31028 Toulouse Cedex 4
\and
	    CNRS, UMR 5187, 9 avenue du Colonel Roche, F-31028 Toulouse Cedex 4
 \and
             Laboratoire d'Astrophysique, Observatoire de Grenoble, 
             BP 53, F-38041 Grenoble Cedex 9
      }

   \date{Received ... ; accepted ...}

 
 \abstract
   {High deuterium fractionation is observed in various types of environment such as prestellar cores, hot cores and hot corinos. It has proven to be an efficient probe to study the physical and chemical conditions of these environments. The study of the deuteration of different molecules helps us to understand their formation. This is especially interesting for complex molecules such as methanol and bigger molecules for which it may allow to differentiate between gas-phase and solid-state formation pathways.}
   {Methanol exhibits a high deuterium fractionation in hot corinos. Since CH$_3$OH is thought to be a precursor of methyl formate  we expect that deuterated methyl formate is produced in such environments. We have searched for the singly-deuterated isotopologue of methyl formate, DCOOCH$_3$, in IRAS 16293-2422, a hot corino well-known for its high degree of methanol deuteration. }
   {We have used the IRAM/JCMT unbiased spectral survey of IRAS 16293-2422 which allows us to search for the DCOOCH$_3$ rotational transitions within the survey spectral range (80-280 GHz, 328-366 GHz). The expected emission of deuterated methyl formate is modelled at LTE and compared with the observations.}
   {We have tentatively detected DCOOCH$_3$ in the protostar IRAS 16293-2422. We assign eight lines detected in the IRAM survey to DCOOCH3. Three of these lines are affected by blending problems and one line is affected by calibration uncertainties, nevertheless the LTE emission model is compatible with the observations.  A simple LTE modelling of the two cores in IRAS 16293-2422, based on a previous interferometric study of HCOOCH$_3$,  allows us to estimate the amount of DCOOCH$_3$ in IRAS 16293-2422.  Adopting an excitation temperature of 100 K and a source size of 2\arcsec and 1\farcs5  for  the A and B cores, respectively,  we find that N$_\mathrm{A, DCOOCH3}$ = N$_\mathrm{B, DCOOCH3}$ $\sim$6 $\times$ 10$^{14}$ cm$^{-2}$. The derived deuterium fractionation is $\sim$ 15 \%, consistent with values for other deuterated species in this source and much greater than that expected from the deuterium cosmic abundance. }
	   { DCOOCH$_3$, if its tentative detection is confirmed, should now be considered in theoretical models that study complex molecule formation and their deuteration mechanisms. Experimental work is also needed to investigate the different chemical routes leading to the formation of deuterated methyl formate.}


\titlerunning{Deuterated methyl formate (DCOOCH$_3$) in IRAS 16293-1422}
\authorrunning{Demyk et al.}
   \maketitle
%

\section{Introduction}

 IRAS 16293-2422 (hereafter IRAS 16293) is a complex source hosting two hot corinos (called "A" and "B") in which many complex organic molecules have been observed: acetonitrile (CH$_3$CN), methyl formate (HCOOCH$_3$), ketene (H$_2$CCO) formic acid (HCOOH), ethanol (C$_2$H$_5$OH), ethyl cyanide (C$_2$H$_5$CN), etc. \citep{cazaux2003,bottinelli2004,bisschop2008}. This source is also characterised by a large degree of deuteration with singly, doubly or triply deuterated molecules. HDCO, CH$_3$OD, CH$_2$DOH,  are detected in IRAS 16293 with a deuterium fractionation of $\sim$ 15 \%, 1.8 \% and 37 \%, respectively, doubly-deuterated formaldehyde (D$_2$CO) and methanol (CHD$_2$OH) are observed with a deuteration of 5 and 7.4 \% \citep{parise2006}. Triply deuterated methanol was also observed in IRAS 16293 with a deuterium fractionation ratio of 1.4 \% \citep{parise2004}. 

These values are much greater than expected from the cosmic D/H ratio ($\sim$ 10$^{-5}$). The process of deuteration is a long-standing and challenging issue. High deuterium fractionation is observed in cold regions such as pre-stellar cores \citep{pillai2007,bacmann2003}, in warmer regions such as hot cores (eg. Orion, \citealt{jacq1993}) and hot corinos  and in photo dissociation regions (PDRs) \citep{leurini2006,pety2007}. The detection of deuterated molecules allows one to study the chemical pathways leading to their formation and to trace the chemical and physical conditions of the observed environments. Whereas in cold environments the deuterium molecular enrichment is explained by gas phase chemistry \citep{roberts2003}, in hot regions it is certainly driven by chemistry on the dust icy surface \citep{tielens1983,ceccarelli2001}. This is well illustrated by the detection of multiply deuterated methanol since methanol cannot be efficiently formed in the gas-phase \citep{turner1998}. Since methanol is thought to be involved in methyl formate formation on grains \citep{bennett2007}, the high deuterium fractionation of CH$_3$OH suggests that deuterated methyl formate should be produced together with methyl formate.  

In this study we present the tentative detection of deuterated methyl formate, DCOOCH$_3$, in IRAS 16293 based on observations from the millimeter/submillimeter spectral survey performed toward the low-mass Class 0 protostar IRAS16293 \citep{caux2010}. The observations are described in Sect. \ref{obs}. The tentative detection of DCOOCH$_3$ is presented in Sect. \ref{res} and its formation is discussed in Sect. \ref{discussion}. 


\section{Observations}
\label{obs}

The  survey was carried out with the IRAM-30m (frequency range 80-280 GHz) and JCMT-15m (frequency range 328-366 GHz) telescopes during the
period January 2004 to August 2006. The angular resolution (HPBW) of the observations varied between 9{\arcsec} and 33{\arcsec}, depending on the used telescope and frequency, and the spectral resolution ranged between 0.3 and 1.2\,MHz.  All observations were performed in Double-Beam-Switch observing mode, with a 90{\arcsec} throw.  Pointing and focus were regularly checked,  the resulting pointing accuracy depends on the weather and the observed frequency but was always better than 5{\arcsec}. The lines assigned to deuterated methyl formate presented in this paper were observed in the 190-270~GHz domain at the IRAM~30m, with a spectral resolution of 1-1.25~MHz and a rms of about 5-15~mK (antenna temperature). For a detailed description of this survey, see the paper presenting the observations and calibrations of the whole survey \citep{caux2010}. Once this paper is published, the data will be made publicly available on the TIMASSS web site (http://www-laog.obs.ujf-grenoble.fr/heberges/timasss/).

\section{Results}
\label{res}

\begin{figure}[!t]
   \includegraphics[scale=.37, angle=90]{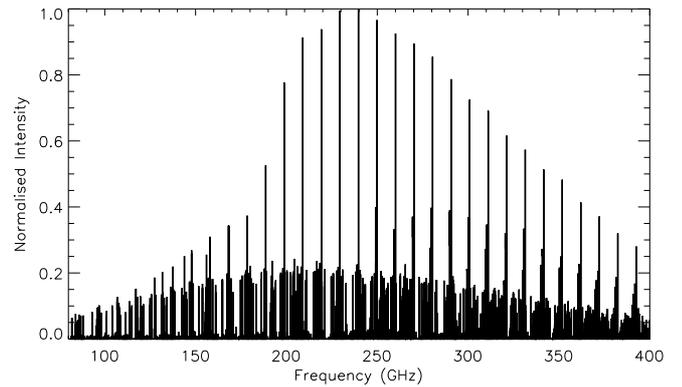}
    \caption{LTE emission model of DCOOCH$_3$ calculated in optically thin conditions for an extended source, an excitation temperature of 100 K, with a linewidth of 5 km.s$^{-1}$. Note that the line intensity has been normalised  to the strongest line at 239.655 GHz.}
    \label{lte}
\end{figure}

To search for deuterated methyl formate we have used the predicted spectrum calculated from the recent study by \citet{margules2010}. Deuterated methyl formate DCOOCH$_3$ is an asymmetric molecule described by the quantum numbers J, K$_{\mathrm{a}}$ and K$_{\mathrm{c}}$. It has a dense  rotational spectrum in which each transition is split into A and E components of the same intensity because of the internal rotation of the methyl group. \\Fig.~\ref{lte} shows the emission spectrum of DCOOCH$_3$ modelled, with the CASSIS software\footnote{The CASSIS software is a free interactive spectrum analyser aiming to interpret astrophysical spectra, for more details see http://cassis.cesr.fr }, in LTE, assuming optically thin conditions, with an excitation temperature of 100 K (typical of hot core values), a line width of 5~km.s$^{-1}$ and an extended source (100\arcsec). The aim of this figure is to investigate the intensity ratios of the DCOOCH$_3$ transitions, and it can indeed be seen that some lines are strikingly more intense than the others. Each of these strong lines consists of 8 transitions with J$^{''}$ $\ge$ 18 and K$_{\mathrm{a}}^{'}$-K$_{\mathrm{a}}^{''}$ = 0-1, 0-0, 1-0 and 1-1 (4 for the A-species and 4 for the E-species, see Table \ref{det-lines-onl}) that are very close in frequency and are blended in the modelled spectrum because of the line width. 

Considering that the line width of complex molecules in IRAS 16293 is of a few km.s$^{-1}$ (2-6 km.s$^{-1}$ for HCOOCH$_3$ \citep{bottinelli2004}), these very intense lines are particularly favourable to search for DCOOCH$_3$. Eleven of these lines are in the observed spectral range of the IRAS 16293 spectral survey, eight in the IRAM data and three in the JCMT data. In the JCMT spectral range, because the modelled line intensity decreases (Fig.~\ref{lte}) and the rms of the observed spectrum is high ($\sim$ 15-25 mK, main beam temperature), the line intensities predicted by the LTE model described below are within the noise. Consequently, we do not detect these transitions. We assign eight lines detected in the IRAM spectral range to the other eight transitions of DCOOCH$_3$ (Table \ref{det-lines-onl}). All other lines from DCOOCH$_3$  are below the detection limit. Indeed, for the other transitions of DCOOCH$_3$, i.e. transitions with J$^{''} \ge$ 18 and K$_{a}^{'} > 1$  and  transitions with J$^{''} < $ 18 and for all value of K$_{a}^{'}$ (i.e.  frequency $<$ 190 GHz) the transition frequencies are more spaced out. Consequently the lines are not blended anymore and their intensity is low (Fig.~\ref{lte}). At 2 and 3 mm, this is combined with beam dilution effect since  the emission region is supposed to be compact.\\

\addtocounter{table}{1} 

\begin{figure*}[th]
\centering
    \includegraphics[width=12.5cm, angle=90]{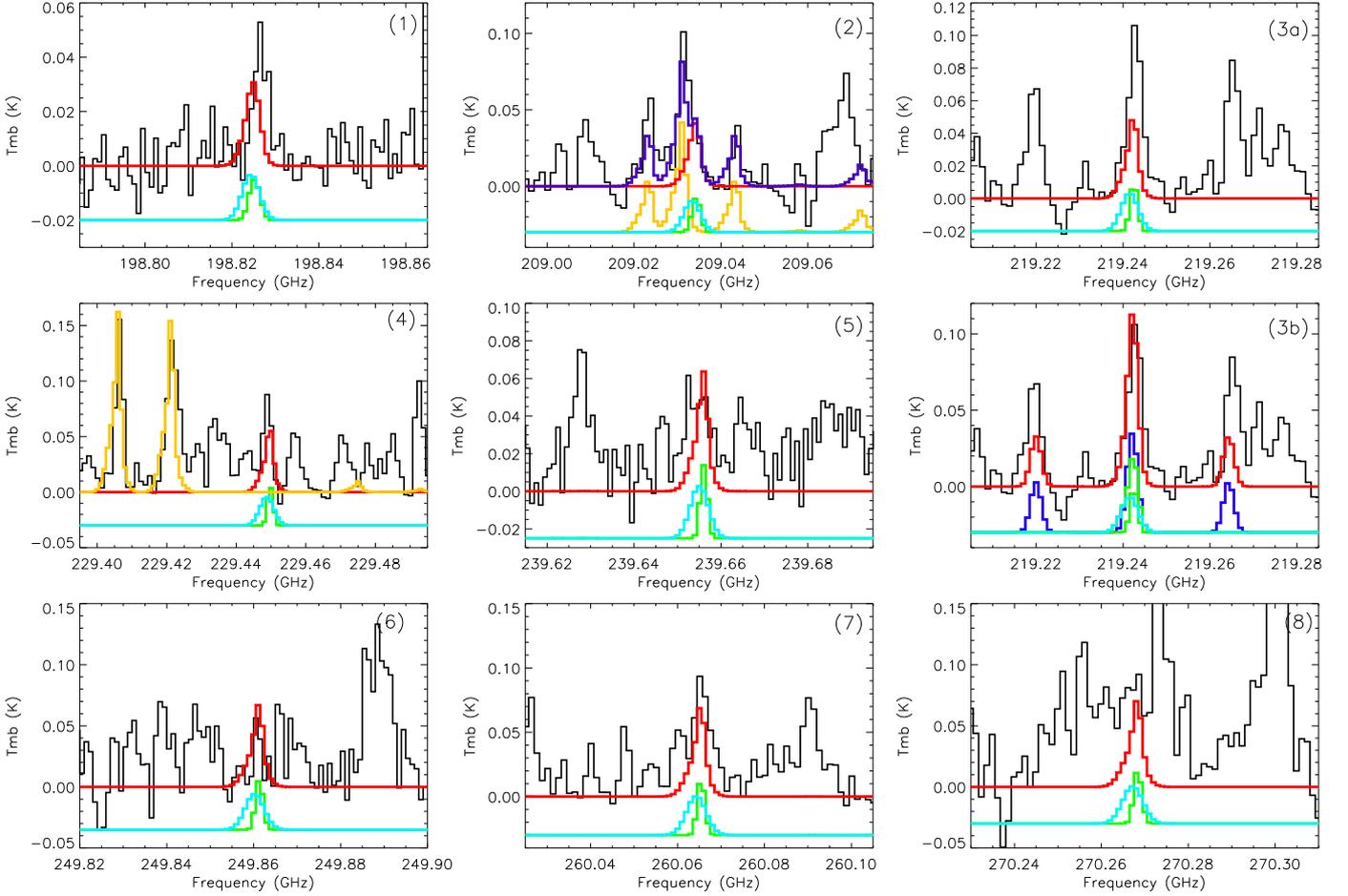}
   \caption{Observed lines assigned to DCOOCH$_3$ in IRAS 16293 compared with a two-components LTE emission model of DCOOCH$_3$. Black line: observed IRAM spectral lines. Red line: sum of the emission from region A (light blue line) modelled with ${\mathrm{\theta}_s}$ = 2\arcsec, T$_\mathrm{ex}$ = 100 K, V$_\mathrm{lsr}$ = 3.9 km.s$^{-1}$ and $\mathrm{\Delta}$v = 6 km.s$^{-1}$, N$_\mathrm{A, DCOOCH3}$ $\sim 6 \times 10^{14}$ cm$^{-2}$ and region B (green line) modelled with  $ {\mathrm{\theta}_s}$ = 1.5\arcsec, T$_\mathrm{ex}$ = 100 K, V$_\mathrm{lsr}$ = 2.7 km.s$^{-1}$ and $\mathrm{\Delta}$v = 2 km.s$^{-1}$, N$_\mathrm{B, DCOOCH3} \sim 6 \times 10^{14}$ cm$^{-2}$. The yellow line in panels (2) and (4) is the emission from HCOOCH$_3$ modelled with the same two components A and B (see text). Panel (3a) and (3b) both show the line at 219.242 GHz: (3a): emission from the two components of DCOOCH$_3$ and their sum, (3b): DCOOCH$_3$ emission (green and light blue lines) and the CH$_3$COCH$_3$ emission (dark blue line, see text), the sum of theses emissions is shown in red. For sake of clarity the light blue,  dark blue and the green lines  have been shifted downward. }
    \label{det}
\end{figure*}

To compare with the observations, we have modelled the DCOOCH$_3$ emission in LTE. Although the single-dish observations do not allow us to constrain the spatial distribution of DCOOCH$_3$ it seems reasonable to assume that it follows that of the main isotopologue. Thus, following the interferometric observations of HCOOCH$_3$ from \citet{bottinelli2004} we have used a two-component emission model characterised as follows: an angular size, $\theta_\mathrm{s}$, of  2\arcsec, V$_\mathrm{lsr}$ = 3.9~km.s$^{-1}$ and a line width $\mathrm{\Delta}$v of 6 km.s$^{-1}$ for source A and  $\theta_\mathrm{s} =$ 1\farcs5, V$_\mathrm{lsr}$ = 2.7~km.s$^{-1}$ and $\mathrm{\Delta}$v = 2 km.s$^{-1}$ for source B. The excitation temperature was set to 100 K for both emission regions, value derived for HCOOCH$_3$ in the spectral survey (see below). A comparison between the observations and the model is presented in Fig. \ref{det}.  \\
The transition at 198.824 GHz is potentially blended with a bright SO line in the image sideband ((5,4)-(4,3) at 206.175 GHz, observed T$\mathrm{_A}^*$=2.5K,  the line shows up at 198.826 GHz in the signal band). However there is a problem with the calibration of the observations at this frequency. Indeed, adopting a sideband rejection of 15 dB (default value), we find that the SO contaminating line from the image band would have T$\mathrm{_A}^* \sim$ 80 mK, in contradiction with the observed line at 198.826 GHz (T$\mathrm{_A}^* \sim$ 35 mK). The assumed sideband rejection seems thus to be incorrect and without knowing the true rejection of this observation we cannot conclude whether the line is due to SO only or due to DCOOCH$_3$ and either SO or another unidentified species. This line should thus be considered with caution. \\
The line at 209.033 GHz is partially blended with a line from HCOOCH$_3$ (A species, 17$_{11,6}$-16$_{11,5}$, 209.030 GHz). Nonetheless, the observations are well reproduced with a model including both the normal and deuterated species (Fig. \ref{det}). To model the HCOOCH$_3$ emission we have used the same two-component LTE model as for the deuterated species. Tens of HCOOCH$_3$ lines are present in the spectral survey. They have been modelled  with an excitation temperature of 100 K and HCOOCH$_3$ abundances of N$_\mathrm{A, HCOOCH3}$ $\sim$1 $\times$ 10$^{16}$ cm$^{-2}$ and N$_\mathrm{B, HCOOCH3}$ $\sim$9 $\times$ 10$^{15}$ cm$^{-2}$. The detailed analysis of this species will be published elsewhere, however these values are consistent with those of \citet{bottinelli2004} within the uncertainties. An excess of emission on the high frequency side of the HCOOCH$_3$ lines at  209.023, 209.033 and  209.044 GHz (although for the latter the line wing is within the noise) (Fig. \ref{det}, panel 2) may be seen in the observed spectrum. Since this excess of emission is attributed to DCOOCH$_3$ in the 209.033 GHz line it is important to assess that it is not related to a HCOOCH$_3$ line wing. This was checked on the numerous spectral lines assigned to HCOOCH$_3$ in the spectral survey. It appears that they do not show any systematic emission excess, as can be seen in the panel (4) of Fig.~\ref{det} which shows two methyl formate lines at 229.405 and 229.420 GHz that do no present such a feature.  \\
The line at 219.242 GHz is partially blended with a strong acetone (CH$_3$COCH$_3$) line at 219.242 GHz (21$_{1,19}$-20$_{2,19}$). Acetone contributes to about two thirds of the intensity of the observed line (N$_\mathrm{CH3COCH3} \sim 5 \times 10^{15}$ cm$^{-2}$, T$_\mathrm{ex}$=100 K$, \theta_\mathrm{s} =$ 1\farcs5, V$_\mathrm{lsr}$ = 2.7~km.s$^{-1}$ and $\mathrm{\Delta}$v = 4 km.s$^{-1}$)  while DCOOCH$_3$ contributes to one third (Fig. \ref{det}, panels 3a and 3b). It should be noted that the intensity of the two acetone lines at 219.220 GHz and 219.265 GHz (Fig. \ref{det}, panel 3b) are not well-reproduced. We have carefully checked that the adopted acetone model does not underestimate the abundance of acetone. It is not possible to account for the whole line integrated intensities of these three lines without being in strong contradiction with other strong acetone lines in the survey. Typically, the intensity of the strongest acetone lines is overestimated by a factor 2 or 3 (depending on the parameters of the model) when the intensity of the 219.220, 219.242 and 219.265 GHz lines are well reproduced. Furthermore, the acetone lines are optically thin, independently of the model parameters (the optical depth of the strongest acetone lines do not exceed ~7$\times$10$^{-2}$). We thus do not expect any opacity problem that could affect the line intensities. The detailed analysis of acetone in the survey is postponed to a future paper. Hence we conclude that these two lines are blended with transitions of other species, although we have not identified with which species, and that the 219.242 GHz band cannot only be due to acetone and is consistent with additional emission from DCOOCH$_3$ .\\
The transitions at 229.446, 239.654, 249. 860 and 260.064 GHz do not appear to be blended. The observed line at 249. 860 GHz is narrower than the modelled line. However we have checked that it is not an instrumental artefact or a spike and it is probably due to the low signal/noise ratio (S/N $\sim$ 1.6).  \\
The spectral range around the line at 270.26 GHz is characterised by an important spectral confusion.  However, although it is contaminated by nearby lines from other (unknown) molecules, the line at 270.267 GHz is reasonably well reproduced by the DCOOCH$_3$ emission model. \\
Thus, among the eight lines in the IRAM spectral range assigned to DCOOCH$_3$, four lines are not blended with other species (229.448, 239.654, 249.860, 260.064 GHz). Two lines are blended with known species (HCOOCH$_3$ at 209.031 GHz  and CH$_3$COCH$_3$ at  219.242 GHz). They are well reproduced by the model and DCOOCH$_3$ contributes to about 30-35 \% to their intensity, independently of the adopted model used for acetone and methyl formate. The line at 270.267 GHz is heavily blended with unknown species but the emission model is in agreement with the observations. The emission model is compatible with the observations at 198.824 GHz, however because this observation suffers from calibration uncertainty, it is not possible to conclude about the detection of this line.\\

For DCOOCH$_3$, a reasonable agreement with the observations is found for a column density of N$_\mathrm{A, DCOOCH3} = \mathrm{N_{B, DCOOCH3}} \sim 6 \times 10^{14}$ cm$^{-2}$. High angular resolution observations are needed not only to confirm the adopted two-component model for DCOOCH$_3$, but also to clarify the situation regarding HCOOCH$_3$. Indeed, in their interferometric study of this species, \citet{bottinelli2004} measured an apparent difference between the V$_\mathrm{lsr}$  and line widths of the two hot corinos, namely V$_\mathrm{lsr}$ = 3.9 km.s$^{-1}$ and $\Delta$v = 6 km.s$^{-1}$ for source A, and V$_\mathrm{lsr}$ = 2.7 km.s$^{-1}$ and $\Delta$v = 2 km.s$^{-1}$ for source B. However the authors were not able to conclude whether the values for the V$_\mathrm{lsr}$  and line width in source B were real or due to the distortion of optically thick HCOOCH$_3$ lines in source B, in which case the blue wing of the (strongly self-absorbed) broad line normally centered at 3.9 km.s$^{-1}$ could be mistaken for a narrow line centered at 2.7 km.s$^{-1}$. The DCOOCH$_3$ lines are optically thin. Therefore, if the two-component model used in this study is confirmed by higher angular resolution observations, it would imply that  the value of V$_\mathrm{lsr}$ and of the line width are indeed different in the two hot corinos in IRAS 16293. \\
Using the column density previously derived for HCOOCH$_3$ and DCOOCH$_3$, the degree of deuteration of methyl formate in IRAS 16293 is estimated to $\sim$ 15\%. This degree of deuteration is similar in source A and B but dedicated high spatial resolution observations are needed to confirm this (HCOOCH$_3$ could be optically thick in the core B). Although this is a rough estimate, the deuterium enrichment of methyl formate, greater than the one expected from deuterium cosmic abundance, is compatible with other singly deuterated species in this source such as HDCO or CH$_2$DOH that have a deuterium enrichment of $\sim$ 15~\% and 37~\%, respectively \citep{parise2006}.

\section{Discussion}
\label{discussion}

It is established that methyl formate cannot be formed exclusively by gas-phase reactions \citep{horn2004} and that solid state reactions necessarily play a role in its formation. Methyl formate may form in the icy mantle of dust grains via the reaction between the metoxy radical, CH$_3$O, with the formyl radical, HCO. CH$_3$O and HCO are synthesized in the grain mantles either by hydrogenation of H$_2$CO and CO, respectively \citep{garrod2006,watanabe2008} or via UV photodissociation of methanol which produces CH$_3$O and an H atom that reacts with CO to produce HCO \citep{bennett2007}. As for methyl formate, it is likely that deuterated methyl formate formation also requires grain-chemistry. We discuss below two formation pathways, implying solid state reactions, that may lead to DCOOCH$_3$. 

The first one is similar to the formation of HCOOCH$_3$. In that scenario, DCOOCH$_3$ is formed directly via the reaction: DCO + CH$_3$O $\rightarrow$ DCOOCH$_3$ (1). At low temperature and in the absence of energetic processes, DCO could be produced by deuteration reaction of CO with gas phase deuterium atom accreted onto the grain while CH$_3$O results from the hydrogenation of H$_2$CO. However, experiments from \citet{hidaka2007} have shown that the deuteration of CO ice is about 10 times smaller than the hydrogenation rate because there is no tunnelling effect in the case of deuteration. The authors thus conclude that the deuteration of CO may not be the first step to produce deuterated formaldehyde or methanol in molecular clouds. This may also be the case for the formation of deuterated methyl formate. \\
Energetic processes in ices, such as UV photolysis or cosmic rays bombardment of ices, may be more efficient to produce DCOOCH$_3$ via reaction (1). Following the formation of methyl formate reported by \citet{bennett2007}, DCO could be formed from the reaction of CO with a deuterium atom produced by the UV photodissociation of deuterated methanol. The energy given by the UV photons should help to overcome the activation energy of the reaction CO + D $\rightarrow$ DCO. Among the deuterated isotopologues of methanol, CH$_3$OD seems to be the best candidate for the UV photodissociation since, if the UV photons break the O-D bond of CH$_3$OD, CH$_3$O and a D atom are easily produced. This is not the case for any other deuterated methanol species, for which the simultaneous production of CH$_3$O and D would require intermediate reactions. 

A second route to form deuterated methyl formate is the H/D substitution in solid HCOOCH$_3$ during the warm-up of the ice mantles. This scenario is proposed for the formation of deuterated formaldehyde, and participates to the formation of deuterated methanol species \citep{hidaka2009,ratajczak2009}. It may also be efficient to produce deuterated methyl formate.
Once again, the H/D substitution could also be driven by energetic processes such as UV photolysis or cosmic rays irradiation of the ice mantles \citep{weber2009}. In their study of methane/deuterated-water ice, \citet{weber2009} concluded that almost all organic species should undergo H/D substitution with the matrix in water ices exposed to UV radiation. 

Isotopic substitutions are routinely used in laboratory experiments to disentangle the different chemical pathways leading to the formation of molecules. From that point of view the formation of deuterated methyl formate is an interesting issue that could improve our understanding of the chemistry of these species. To our knowledge, experimental and theoretical works on the deuteration of methyl formate and other complex organic molecules do not exist yet. They are clearly needed to distinguish between the above scenarii.  The astronomical detection of other deuterated isotopologues of methyl formate  should also put some constraints on the different formation pathways.

\section{Conclusion}
We have tentatively detected the singly-deuterated isotopologue of methyl formate, DCOOCH$_3$, in the protostar IRAS16293. We have assigned eight observed lines to DCOOCH$_3$ transitions. Four lines are not blended with other species (229.448, 239.654, 249.860, 260.064 GHz). Three lines are blended (209.031, 219.242, 270.267 GHz) with other species. Among the latter, the first two lines are blended with HCOOCH$_3$ and CH$_3$COCH$_3$, respectively. They are well-reproduced by an emission model at LTE. We were not able to identify the species responsible for the blending of the 270.267 GHz line however the LTE model is compatible with  the observations and does not contradict the tentative identification of DCOOCH$_3$ in IRAS 16293. The LTE emission model is compatible with the last observed line at 198.824 GHz but we cannot conclude on its detection because of calibration uncertainty. From a basic modelling in LTE we estimate the abundance of deuterated methyl formate to be N$_\mathrm{A+B, DCOOCH3}$ $\sim$1.2 $\times$ 10$^{15}$ cm$^{-2}$. This leads to a DCOOCH$_3$/HCOOCH$_3$ ratio of  $\sim$ 15 \%, consistent with the deuteration fractionation of other singly deuterated species in this source.

Additional observations with better spectral resolution and higher sensitivity are needed for several reasons. First, considering the low spectral resolution and the relatively small signal-to-noise ratio of the lines assigned to DCOOCH$_3$, such new observations would strengthen this tentative detection. Second, it would allow to detect additional lines among the hundreds weaker lines from DCOOCH$_3$ that are present in the spectral survey. Lastly, it would be the opportunity to search for other deuterated isotopologues of methyl formate. High angular resolution observations would also be helpful to investigate the existence of two emission sources for DCOOCH$_3$ as it is observed for HCOOCH$_3$ and to conclude about the reality of the physical differences between the two cores A and B. Such observations are also necessary to properly estimate the methyl formate deuteration and, together with new experimental and theoretical work, to understand the deuteration mechanisms of complex organic molecules.

\begin{acknowledgements}
The authors would like to thank Laurent Margul\`es, Miguel Carvajal and Isabelle Kleiner for providing us with the DCOOCH$_3$ spectral data prior to publication. We thank the IRAM staff for help provided during the observations. Support by the French National Agency (ANR-02-BLAN-0225-01) is acknowledged.
\end{acknowledgements}

\bibliographystyle{aa}
\bibliography{bib.bib}

\longtab{1}{  \begin{longtable}{ c c c c c c c c c }
\caption{ Emission lines assigned to DCOOCH$_3$ in the IRAS 16293 IRAM spectral survey.} \label{det-lines-onl} \\ 
\hline \hline
Trans. & Pred.  & S${\mu}^2$    &   E$_{\mathrm{up}}$ & Obs.             & Obs.  					     &  Uncertertainty                                    & Pred.  					 \\
$ \mathrm{ J^{'}_{ K^{'}_{a}K^{'}_{c} } - J^{''}_{ K^{''}_{a}K^{''}_{c} } } $   &   Freq.       &        &       & Freq.\tablefootmark{a}    &  $\int {\mathrm{T_{mb}\delta v}} $ &  Obs. $\int {\mathrm{T_{mb}\delta v}}$  & $\int {\mathrm{T_{mb}\delta v}} $ \\
         & (MHz) & (D$^2$)          &  (cm$^{-1}$)  	     &  (MHz)          &  (mK. km.s$^{-1}$) \tablefootmark{b}             & (mK. km.s$^{-1}$) \tablefootmark{c}                       & (mK. km.s$^{-1}$) \tablefootmark{d}     	  \\
\endfirsthead
\caption{Emission lines assigned to DCOOCH$_3$  in the IRAS 16293 IRAM spectral survey -- continued from previous page.} \\
\hline
\hline
Trans. & Pred.  & S${\mu}^2$    &   E$_{\mathrm{up}}$ & Obs.             & Obs.  					     &  Uncertertainty                                    & Pred.  					 \\
$ \mathrm{ J^{'}_{ K^{'}_{a}K^{'}_{c} } - J^{''}_{ K^{''}_{a}K^{''}_{c} } } $   &   Freq.       &        &       & Freq.\tablefootmark{a}    &  $\int {\mathrm{T_{mb}\delta v}} $ &  Obs. $\int {\mathrm{T_{mb}\delta v}}$  & $\int {\mathrm{T_{mb}\delta v}} $ \\
         & (MHz) & (D$^2$)          &  (cm$^{-1}$)  	     &  (MHz)          &  (mK. km.s$^{-1}$) \tablefootmark{b}             & (mK. km.s$^{-1}$) \tablefootmark{c}                       & (mK. km.s$^{-1}$) \tablefootmark{d}     	  \\
\hline
\endhead
\hline 
\multicolumn{8}{r}{\it{continued on next page}} \\
\endfoot
\hline 
\hline 
\endlastfoot
\hline 
19$_{0,19}$-18$_{1,18}$ E &  198\,821.94 &  \multirow{8}{*}{ $ \left.\begin{array}{r} 
  9.77 \\  
  9.77 \\
 50.85 \\
 50.86 \\
 50.85 \\
 50.86 \\
  9.77 \\
  9.76  
\end{array}\right\} $ } & \multirow{8}{*}{61} & \multirow{8}{*}{198\,826.65$^*$ } & \multirow{8}{*}{289} & \multirow{8}{*}{63} &\multirow{8}{*}{258}	 \\
19$_{0,19}$-18$_{1,18}$ A &  198\,823.01 &  &  &  \\
19$_{1,19}$-18$_{1,18}$ E &  198\,823.21 &  &  &  \\
19$_{1,19}$-18$_{1,18}$ A &  198\,824.28 &  &  &  \\
19$_{0,19}$-18$_{0,18}$ E &  198\,824.38 &  &  &  \\
19$_{0,19}$-18$_{0,18}$ A &  198\,825.45 &  &  &  \\
19$_{1,19}$-18$_{0,18}$ E &  198\,825.65 &  &  &  \\
19$_{1,19}$-18$_{0,18}$ A &  198\,826.72 &  &  &  \\
20$_{0,20}$-19$_{1,19}$ E &  209\,031.87 & \multirow{8}{*}{ $ \left.\begin{array}{r} 
10.32 \\
53.55 \\
10.32 \\
53.55 \\
53.56 \\
10.32 \\
10.32 \\
53.56
\end{array}\right\} $ }  & \multirow{8}{*}{68} & \multirow{8}{*}{209\,031.80$^*$ } &  \multirow{8}{*}{546} &  \multirow{8}{*}{59} &  \multirow{8}{*}{275}\\
20$_{1,20}$-19$_{1,19}$ E &  209\,032.54 &   &  &  \\
20$_{0,20}$-19$_{1,19}$ A &  209\,032.89 &   &  &  \\
20$_{0,20}$-19$_{0,19}$ E &  209\,033.15 &   &  &  \\
20$_{1,20}$-19$_{1,19}$ A &  209\,033.55 &   &  &  \\
20$_{1,20}$-19$_{0,19}$ E &  209\,033.81 &   &  &  \\
20$_{0,20}$-19$_{0,19}$ A &  209\,034.16 &   &  &  \\
20$_{1,20}$-19$_{0,19}$ A &  209\,034.82 &   &  &  \\
21$_{0,21}$-20$_{1,20}$ E &  219\,240.53 & \multirow{8}{*}{ $ \left.\begin{array}{r} 
10.88 \\
56.26 \\
56.26 \\
10.87 \\
10.88 \\
56.26 \\
56.26 \\
10.87
\end{array}\right\} $ }  & \multirow{8}{*}{75} & \multirow{8}{*}{219\,242.75$^*$ } &  \multirow{8}{*}{614} &  \multirow{8}{*}{54} &  \multirow{8}{*}{281} \\
21$_{1,21}$-20$_{1,20}$ E &  219\,240.88 &   &  &  \\
21$_{0,21}$-20$_{0,20}$ E &  219\,241.20 &   &  &  \\
21$_{0,21}$-20$_{1,20}$ A &  219\,241.49 &   &  &  \\
21$_{1,21}$-20$_{0,20}$ A &  219\,241.54 &   &  &  \\
21$_{1,21}$-20$_{1,20}$ A &  219\,241.84 &   &  &  \\
21$_{0,21}$-20$_{0,20}$ A &  219\,242.16 &   &  &  \\
21$_{1,21}$-20$_{0,20}$ A &  219\,242.50 &   &  &  \\
22$_{0,22}$-21$_{1,21}$ E &  229\,448.05 & \multirow{8}{*}{ $ \left.\begin{array}{r} 
11.43 \\
58.96 \\
58.96 \\
11.43 \\
11.43 \\
58.97 \\
58.97 \\
11.43
\end{array}\right\} $ }  & \multirow{8}{*}{82} &  \multirow{8}{*}{229\,448.88} &  \multirow{8}{*}{297} 	&  \multirow{8}{*}{89} &  \multirow{8}{*}{275} \\
22$_{1,22}$-21$_{1,21}$ E &  229\,448.23 &   &  &  \\
22$_{0,22}$-21$_{0,21}$ E &  229\,448.40 &   &  &  \\
22$_{1,22}$-21$_{0,21}$ E &  229\,448.57 &   &  &  \\
22$_{0,22}$-21$_{1,21}$ A &  229\,448.95 &   &  &  \\
22$_{1,22}$-21$_{1,21}$ A &  229\,449.13 &   &  &  \\
22$_{0,22}$-21$_{0,21}$ A &  229\,449.29 &   &  &  \\
22$_{1,22}$-21$_{0,21}$ A &  229\,449.47 &   &  &  \\
23$_{0,23}$-22$_{1,22}$ E &  239\,654.45 & \multirow{8}{*}{ $ \left.\begin{array}{r} 
11.99 \\
61.66 \\
61.66 \\
11.99 \\
11.98 \\
61.67 \\
61.67 \\
11.98
\end{array}\right\} $ }  & \multirow{8}{*}{90} & \multirow{8}{*}{239\,654.07} & \multirow{8}{*}{401}  & \multirow{8}{*}{103} &  \multirow{8}{*}{291}	\\
23$_{1,23}$-22$_{1,22}$ E &  239\,654.55 &   &  &  \\
23$_{0,23}$-22$_{0,22}$ E &  239\,654.63 &   &  &  \\
23$_{1,23}$-22$_{0,22}$ E &  239\,654.72 &   &  &  \\
23$_{0,23}$-22$_{1,22}$ A &  239\,655.30 &   &  &  \\
23$_{1,23}$-22$_{1,22}$ A &  239\,655.39 &   &  &  \\
23$_{0,23}$-22$_{0,22}$ A &  239\,655.47 &   &  &  \\
23$_{1,23}$-22$_{0,22}$ A &  239\,655.57 &   &  &  \\
24$_{0,24}$-23$_{1,23}$ E &  249\,859.75 & \multirow{8}{*}{ $ \left.\begin{array}{r} 
12.54 \\
64.37 \\
64.37 \\
12.54 \\
12.54 \\
64.38 \\
64.38 \\
12.54
\end{array}\right\} $ }  & \multirow{8}{*}{98} & \multirow{8}{*}{249860.81} &  \multirow{8}{*}{107}   &  \multirow{8}{*}{68} &  \multirow{8}{*}{303} \\
24$_{1,24}$-23$_{1,23}$ E &  249\,859.79 &   &  &  \\
24$_{0,24}$-23$_{0,23}$ E &  249\,859.84 &   &  &  \\
24$_{1,24}$-23$_{0,23}$ E &  249\,859.89 &   &  &  \\
24$_{0,24}$-23$_{1,23}$ A &  249\,860.54 &   &  &  \\
24$_{1,24}$-23$_{1,23}$ A &  249\,860.59 &   &  &  \\
24$_{0,24}$-23$_{0,23}$ A &  249\,860.63 &   &  &  \\
24$_{1,24}$-23$_{0,23}$ A &  249\,860.68 &   &  &  \\
25$_{0,25}$-24$_{1,24}$ E &  260\,063.90 &  \multirow{8}{*}{ $ \left.\begin{array}{r} 
13.10 \\
67.07 \\
67.07 \\
13.10 \\
13.10 \\
67.09 \\
67.09 \\
13.10
\end{array}\right\} $ }  & \multirow{8}{*}{106} & \multirow{8}{*}{260\,064.61} &  \multirow{8}{*}{574}  &  \multirow{8}{*}{61} &  \multirow{8}{*}{308} \\ 
25$_{1,25}$-24$_{1,24}$ E &  260\,063.92 &   &  &  \\
25$_{0,25}$-24$_{0,24}$ E &  260\,063.94 &   &  &  \\
25$_{1,25}$-24$_{0,24}$ E &  260\,063.97 &   &  &  \\
25$_{0,25}$-24$_{1,24}$ A &  260\,064.63 &   &  &  \\
25$_{1,25}$-24$_{1,24}$ A &  260\,064.66 &   &  &  \\
25$_{0,25}$-24$_{0,24}$ A &  260\,064.68 &   &  &  \\
25$_{1,25}$-24$_{0,24}$ A &  260\,064.71 &   &  &  \\
\newpage
26$_{0,26}$-25$_{0,25}$ E & 270\,266.86 &  \multirow{8}{*}{ $ \left.\begin{array}{r} 
13.65 \\ 
69.79 \\
69.79 \\
13.65 \\
13.64 \\
69.79 \\
69.79 \\
13.64
\end{array}\right\} $ }  & \multirow{8}{*}{115} & \multirow{8}{*}{270\,267.24$^*$ } &  \multirow{8}{*}{-}  &  \multirow{8}{*}{ 15 mK } &  \multirow{8}{*}{363} \\
26$_{1,26}$-25$_{1,25}$ E & 270\,266.88 &  &  &  \\
26$_{0,26}$-25$_{1,25}$ E & 270\,266.89 &  &  &  \\
26$_{1,26}$-25$_{0,25}$ E & 270\,266.90 &  &  &  \\
26$_{0,26}$-25$_{1,25}$ A & 270\,267.55 &  &  &  \\
26$_{0,26}$-25$_{0,25}$ A & 270\,267.56 &  &  &  \\
26$_{1,26}$-25$_{1,25}$ A & 270\,267.57 &  &  &  \\
26$_{1,26}$-25$_{0,25}$ A & 270\,267.58 &  &  &  \\
\end{longtable}
\tablefoot{ The lines with an asterisk are blended with other species, see text.\\
\tablefoottext{a}{Observed (centroid) frequencies assuming that the radial velocities relative to LSR are 3.9 km s$^{-1}$, it does not coincide with the predicted frequency because it does not take into account blending with transitions of other species.}  
\tablefoottext{b}{Observed integrated intensity in main-beam temperature scale.  Because of the high spectral confusion level it was not calculated for the line at 270.267 GHz.}  
\tablefoottext{c}{Uncertainty on the observed integrated intensity in main-beam temperature scale calculated as follow: $\sigma = (1+\alpha) \times$ rms $\times {\delta}$v $\times \sqrt{\mathrm{N}}$, N being the number of channels over which the integrated intensities are computed, $\delta$v the channel width and $\alpha$ the uncertainty on the calibration of the observations ($\alpha$ = 17\% in the 198-265 GHz spectral range, see \citet{caux2010}). For the line at 270.267 GHz, the rms of the observed spectrum in main-beam temperature scale is given.}  
\tablefoottext{d}{Integrated intensity of the modelled lines.}
}
}

\end{document}